\documentclass[11pt]{amsart}

\usepackage{graphicx}
\usepackage{hyperref}
\usepackage{url}

\input{pstricks}
\input{pst-node}
\usepackage{pst-tree}

\newcommand{\tree}[2]{\pstree[treemode=U,arrows=->,treefit=tight,treesep=0.5cm,levelsep=1cm]{#1}{#2}}
\newcommand{\node}[1]{\Tr{\psframebox[linecolor=white,framearc=.5]{#1}}}
\renewcommand{\root}[1]{\Toval{#1}}

\DeclareMathOperator{\rank}{rank}

\DeclareMathOperator{\diag}{diag}

\newcommand{\ind}{\mbox{$\perp \kern-5.5pt \perp$}}

\newcommand{\pa}{\mathrm{pa}}  
\newcommand{\cE}{\mathcal{E}}
\newcommand{\cG}{\mathcal{G}}

\newcommand{\RP}{\mathcal R}  

\newcommand{\bU}{\mathbf U}
\newcommand{\bI}{\mathbf I}
\newcommand{\muta}[3]{\rho_{#1,#2}^{#3}}
\newcommand{\thet}[1]{\theta^e_{#1_{\pa(e)}, #1_e}}

\newtheorem{thm}{Theorem}

\newtheorem{prop}[thm]{Proposition}
\newtheorem{cor}[thm]{Corollary}

\newtheorem{ex}[thm]{Example}

\title{Evolution on distributive lattices}

\author[Beerenwinkel, Eriksson, and Sturmfels]{
Niko Beerenwinkel$^*$ \and Nicholas Eriksson  \and Bernd Sturmfels\\
Department of Mathematics\\
University of California\\
Berkeley, CA 94720, USA\\
$\{$niko,eriksson,bernd$\}$@math.berkeley.edu\\
$^*$Corresponding Author:\\
phone: +1 (510) 642-3529, fax: +1 (510) 642-8204
}

\begin{document}

\begin{abstract}
We consider the directed evolution of a population after an 
intervention that has significantly altered the underlying 
fitness landscape. 
We model the space of genotypes as a distributive lattice;
the fitness landscape is a real-valued function on 
that lattice. The risk of escape from intervention, i.e., the 
probability that the population
develops an escape mutant before extinction, is
encoded in the  risk polynomial.
Tools from algebraic combinatorics are applied
to compute the risk polynomial in terms of
the fitness landscape. In an application to 
 the development of drug
resistance in HIV, we study the
 risk of viral escape from
treatment with the protease inhibitors ritonavir
and indinavir.     
\end{abstract}

\maketitle

\begin{quote}
\end{quote}

\section{Introduction}

The evolutionary fate of a population is determined by the replication
dynamics of the ensemble and by the reproductive success of its individuals.
We are interested in scenarios where most individuals have a low fitness,
eventually leading to extinction, and only a few types of individuals
(``escape mutants'')
can survive permanently. These situations often arise due to  
a significant change of the underlying fitness landscape. 
For example, a virus
that has been transmitted to a new host is confronted with a new immune
response. Likewise, medical interventions such as radiation therapy,
vaccination, or chemotherapy result in altered fitness landscapes for the 
targeted agents, which may be bacteria, viruses, or cancer cells.

Given a population and such a hostile fitness landscape, the central question
is whether the population will survive.
In the case of medical interventions we wish to know the probability
of successful treatment. Answering this question involves computing
the risk of evolutionary escape, i.e., the probability that the
population develops an escape mutant before extinction.     
We present a mathematical framework for computing such probabilities.

Our primary application is the evolution of drug resistance
during treatment of HIV infected patients \cite{Clavel2004}.
We consider therapy with two different protease inhibitors (PIs).
These compounds interfere with HIV particle maturation
by inhibiting the viral protease enzyme.
The effectiveness of PI therapy is limited
by the development of drug resistance.
Rapid and highly error prone replication of a large virus
population generates mutants that resist the selective pressure of
drug therapy. PI resistance is caused by mutations in the protease gene
that reduce the binding affinity of the drug to the enzyme.
These mutations have been shown to accumulate in a stepwise manner
\cite{Berkhout1999}. For most PIs, no single mutation confers
a significant level of resistance, but multiple mutations are
required for escape from drug pressure.
Quantitative predictions of the probability of successful PI treatment
would help in finding effective antiretroviral
combination therapies. Selecting a drug combination
amounts to controlling the viral fitness landscape.

We regard the directed evolution of a population towards an escape state
as a fluctuation on a fitness landscape. The space of
genotypes is modeled as follows. We start with a 
finite partially ordered set (poset) $\cE$ whose elements are called 
\emph{events}. The events are non-reversible
mutations with some constraints on their order of occurrence.
Such constraints are primarily due to
epistatic effects between different loci in a genome
\cite{Bonhoeffer2000}.
The event constraints define the poset structure: 
$\,e_1 < e_2 \,$ in $\cE$ means that
event $e_1$ must occur before event $e_2$ can occur.
Each genotype $g$ is represented by a subset of $\cE$, namely,
the set of all events that occurred to create $g$. 
Thus a genotype $g$ is an \emph{order ideal} in the  poset $\cE$.
The space of genotypes $\cG$ is the set of
all order ideals in $\cE$, which is a {\em distributive lattice}
\cite[Sec.~3.4]{Stanley1999}.
The order relation on $\cG$ is set inclusion and
corresponds to the accumulation of mutations. 
This mathematical formulation is reasonable in the above situations,
where a population is exposed to strong selective pressure.  

\begin{figure}
\includegraphics[width=\textwidth]{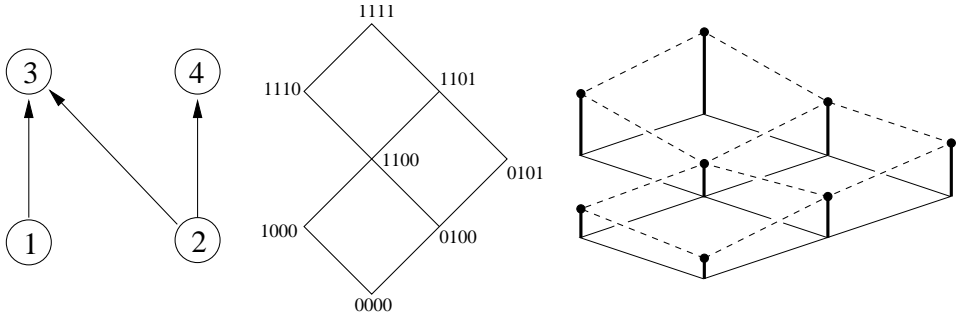}
\caption{An event poset, its genotype lattice, and a fitness landscape.}
\label{fig:ex1}
\end{figure}

The risk of escape is governed by the structure of $\cG$,
the fitness function on $\cG$, and the population dynamics
(such as the mutation rates and population size). Our focus
is on the dependency of the risk of escape
on the assigned fitness values for each genotype $g \in \cG$.
This leads us to the \emph{risk polynomial},
which is shown to be equivalent to a well-known object in
algebraic combinatorics. Indeed, one of the objectives of this
work is to provide a bridge between algebraic combinatorics 
and evolutionary biology.

This paper  is organized as follows. In Section~\ref{sec:fitness}
we formalize our
model of a static fitness landscape on the genotype lattice $\cG$
derived from an event poset $\cE$,
and we discuss evolution on the lattice $\cG$. 
In Section~\ref{sec:branching} we review the multistate
branching process studied by Iwasa, Michor and Nowak 
\cite{Iwasa2003,Iwasa2004}.

In Section~\ref{sec:bayes} we study the Bayesian networks 
which arise from identifying the events in $\cE$
with binary random variables. These
statistical models can be used
to infer the genotype space from
given data. For conjunctive Bayesian networks
we recover the distributive lattice of order ideals in $\cE$.
Of particular interest is
the case where $\cE$ is a directed forest: here the Bayesian network
is a mutagenetic tree model \cite{Beerenwinkel2005c,Beerenwinkel2005f}.
The application of our methods
to the development of PI resistance in HIV 
is presented in  Section~\ref{sec:apply}.

The Appendix summarizes various representations
of the risk polynomial in terms of structures from
algebraic combinatorics. Efficient methods for computing
the risk polynomial and their implementation are presented.


\section{Fitness landscapes on distributive lattices}   \label{sec:fitness}

A partially ordered set (or poset) is a set $\cE$
together with a binary relation, denoted ``$\leq$'', which is
reflexive, antisymmetric, and transitive. Here
we fix a finite poset $\cE$  whose elements are called \emph{events}.
If the number of events is $n$ then we
often identify the set underlying $\cE$ with
the set $\,[n] = \{1,2,\ldots,n\}$. In this way,
the subsets of $\cE$ are encoded by the $2^n$ binary strings of length $n$.
The empty subset of $\cE$ is
encoded by the all-zero string $\hat{0} = 0 0 \cdots 0$ 
which represents the \emph{wild type}, and
the full set $\cE$ is
encoded by the  all-one string $\hat{1} = 11  \cdots 1$
which represents the \emph{escape state}.

An order ideal $g$ in a poset $\cE$ is a subset of $\cE$ 
that is closed downward;
that is, if $e_2 \in g$ and $e_1 \le e_2$, then $e_1 \in g$. 
The set of all order ideals of $\cE$ forms a distributive lattice 
$J(\cE)$ under inclusion. Birkhoff's Representation Theorem
\cite[Thm.~3.4.1]{Stanley1999}
states that all distributive lattices have the form 
$J(\cE)$ for a poset $\cE$.
We write $\cG = J(\cE)$, and we
call $\cG$ the {\em genotype lattice}.

\begin{ex} \rm
Let $\cE$ be the trivial poset, 
where no two events are comparable,
with $|\cE| = n$.
Then $\cG = J(\cE)$ is the Boolean lattice consisting
of all subsets of $\cE$ ordered by inclusion.
This means that all possible combinations of mutations
are possible, and they can occur in any order. Each of
the $2^n$ binary strings $g \in \{0,1\}^n$ represents
a mutational pattern, or genotype.
\end{ex}

In general, the event poset $\cE$ does have non-trivial
relations $e_1 < e_2$. The relation $e_1 < e_2$ excludes all
genotypes $g$ with $g_{e_1} = 0$ and 
$g_{e_2} = 1$ from $\cG$. The remaining genotypes
$g$ form a sublattice of the Boolean lattice $\{0,1\}^n$,
and this is precisely our distributive lattice
$\cG = J(\cE)$. Note that the
lattice $\cG$ is ranked, with the rank function given by
$\rank(g) = |g|$.

\begin{ex} \rm \label{FourEvents}
Consider a scenario with $n=4$ mutation events, labeled $\cE = \{1,2,3,4\}$. 
Suppose that event $3$ can only  occur after
events $1$ and $2$,
and event $4$ can only occur after event $2$.
This allows for precisely eight genotypes
\[
\cG \,\, = \,\, \bigl\{
0000, 1000, 0100, 1100, 0101, 1110, 1101, 1111  \bigr\}.
\]
The event poset $\cE$ and the genotype lattice $\cG$ are
shown in Figure~\ref{fig:ex1}.
\end{ex}

A fitness landscape associates to each possible genotype
a number which quantifies the reproductive capacity of
an individual with that genotype \cite{Reidys2002}. We define a
\emph{fitness landscape} on the distributive lattice $\cG$ 
 to be any function ${\mathbf f} \colon \cG \to \mathbb{R}$.
The value  ${\mathbf f}(g)$ at any $g \in \cG$ 
is the  \emph{fitness} of the genotype $g$.
Thus, the space of all fitness landscapes is the finite-dimensional
vector space $\mathbb{R}^\cG$.

We shall consider certain special models of fitness landscapes,
which are represented by linear subspaces of $\mathbb{R}^\cG$.
In the following definitions, a genotype $g$ is regarded
as a subset of the event poset $\cE$, where $|\cE| = n$.
A \emph{constant fitness landscape} has the
form ${\mathbf f}(g) \equiv a$ for some constant $a$.
Thus the constant landscapes form a
line through the origin in $\mathbb{R}^\cG$.
A \emph{graded fitness landscape} is a landscape on 
$\cG$ whose fitness values depend only on the rank. Equivalently, we have
${\mathbf f}(g) = a_{|g|}$ for
constants $a_0,a_1,\ldots,a_n$. Thus, graded fitness landscapes
form an $(n+1)$-dimensional linear subspace of $\mathbb{R}^\cG$.

Our biological application in Section~\ref{sec:apply} uses
the graded fitness landscape model, which means that the
fitness of a virus type depends only on the number of mutations it
harbors. We shall  
model situations where a virus escapes from a wild
type $\hat{0}$ to a drug-resistant type $\hat{1}$.  In this case, we
assume a graded fitness landscape that is
monotonically increasing with rank, i.e.,
\[ 
   a_0 \,<\, a_1 \,< \,a_2 \,<\, \cdots \,<\, a_n. 
\]
  This implies that the fitness landscape ${\mathbf f}$ has a unique
local (and global) maximum at the drug resistant type $\hat{1}$,
which is the top element in $\cG$.

We next introduce the mathematical framework
for evolution on a fitness landscape. The general
setup is as in the work of Reidys and Stadler
\cite{Reidys2002}, but this is adapted here to our specific
situation, where the genotypes form a 
 distributive lattice $\cG$. The order relation on $\cG$,
 which comes from inclusion of subsets of $\cE$,  induces a
neighborhood structure on $\cG$ where the neighbors
of $g \in \cG$ are the genotypes that strictly contain $g$,
\begin{equation}   \label{eq:neighborhood}
   N(g) \, := \, \bigl\{ h \in \cG \,\mid \, g \subset h \bigr\}.
\end{equation}
Unlike the typical situation considered in \cite{Reidys2002},
this notion of neighborhood is not symmetric. To be precise,
we have that $h \in N(g)$ implies $g \not\in N(h)$.

This neighborhood structure implies that mutational 
changes are possible only upward in the genotype lattice. 
This structure models a directed evolutionary 
process from the wild type $\hat{0}$ towards the escape state
$\hat{1}$. Typically, our configuration space $\cG$ is a small subset
of the Boolean lattice $\{0,1\}^n$ of all binary strings.
Indeed, in the course of viral evolution,
 a population will visit only a small fraction of $\{0,1\}^n$,
 as most mutants are not viable.

Suppose that the number of genotypes in $\cG$ is $m$.
 We wish to define dynamics between the states of $\cG$.
 To this end, we fix a linear extension of $\cG$, and we
  introduce an
 $m \times m$ matrix of transition rates, written
 ${\bf U} = (u_{gh})$, whose rows and columns
 are indexed by genotypes $g,h \in \cG$.
Each entry $u_{gh}$ of the matrix ${\bf U}$ is a non-negative 
real number which is zero unless $h \in N(g)$.
In the framework of algebraic combinatorics, it
is convenient to think of the matrix ${\bf U}$ as an element in the
incidence algebra of $\cG$;
see \cite[Sec.~3.6]{Stanley1999}.

We further assume that the non-zero mutation rates
$u_{gh}$ depend only on the events in $h \backslash g$.
Equivalently, the rate at which a collection of mutation events
occurs is independent of which other mutations have
already occurred. With this assumption, there are only $n$ free
parameters $\mu_1,\ldots,\mu_n$ in the matrix ${\bf U}$,
where $\mu_e$ is the mutation rate of event $e$.
Then 
\begin{equation} \label{eq:muta}
u_{gh} \,\,=\,\, \begin{cases}
   \,\,\,  \prod_{e \in h\backslash g} \mu_e & \text{if $g \subset h$}\\
   \,\,\,  0                               & \text{otherwise}.
   \end{cases} 
\end{equation} 
In particular, if all rates are the same, say $\mu = \mu_1 = \dots = \mu_n$, then
the entries of $\bU$ are $\, u_{gh} \,=\,  \mu^{|h \backslash g|}\,$
if $g \subset h$ and $\,u_{gh} = 0\,$ otherwise.

\begin{ex} \rm \label{FourEvents2}
For the genotype lattice $\cG$ in
Figure~\ref{fig:ex1}, the matrix $\bU$ equals
\[
\bordermatrix{ & \! 0000 \! &
\! 1000 \! & \! 0100 \! & \! 1100 \! & \! 0101 \! & 1110 & 1101 & 1111 \cr
0000 &           0  & \mu_1 & \mu_2 & \mu_1 \mu_2 & \mu_2 \mu_4 & 
\! \mu_1 \mu_2 \mu_3 \! & \! \mu_1 \mu_2 \mu_4 \!&
\! \mu_1 \mu_2 \mu_3 \mu_4 \! \cr
1000 & 0 & 0 & 0 &\mu_2 & 0 & \mu_2 \mu_3 & \mu_2 \mu_4 & \mu_2 \mu_3 \mu_4 \cr
0100 & 0 & 0 & 0 &\mu_1 & \mu_4 & \mu_1 \mu_3 & \mu_1 \mu_4 & \mu_1 \mu_3 \mu_4
 \cr
1100 &        0    &  0   &  0   &  0   &  0   & \mu_3  & \mu_4 & \mu_3 \mu_4  
\cr
0101 &        0    &  0   &  0   &  0   &  0   &  0  &  \mu_1 & \mu_1 \mu_3 \cr
1110 &        0    &  0   &  0   &  0   &  0   &  0  &  0   &  \mu_4   \cr
1101 &        0    &  0   &  0   &  0   &  0   &  0  &  0   &  \mu_3   \cr
1111 &        0    &  0   &  0   &  0   &  0   &  0  &  0   &  0   \cr}
\]
Note that the entry in row $g$ and column $h$ of
any power $\bU^k$ equals $u_{gh}$ times the number
of paths of length $k$ from $g$ to $h$ in $\cG$. In particular,
$\,\bU^5 = 0 $.
\end{ex}

Let ${\mathbf f}$ be a fitness landscape on $\cG$ and  $\,{\mathbf F} \,=\, 
\diag\bigl({\bf f}(g) \mid g \in \cG \bigr)\,$  the $m \times m$ diagonal 
matrix whose entries are the fitness values.
The entry of the matrix product ${\bU} {\mathbf F}$ in row $g$ and column $h$
represents the  probability of genotype $g$ transitioning
into genotype $h$ in one step.
A precise probabilistic derivation and interpretation 
will be given in the next section.

We are interested
in \emph{all} mutational pathways that lead from the wild type
$\hat{0}$ to the escape state $\hat{1}$.
Towards this end, note that the entry $(g,h)$ of the matrix
$({\bU}{\mathbf F})^k$ represents the probability of
genotype $g$ evolving to genotype $h$
along any mutational pathway (chain) of length $k$ in 
the genotype lattice $\cG$.
The chains from $\hat{0}$ to $\hat{1}$ in $\cG$ 
are accounted for by the 
upper right hand entry of $({\bU}{\mathbf F})^k$.
Note that the matrix $\,({\bU}{\mathbf F})^k\,$ is zero for $k > n$.

To account for chains of arbitrary length, we consider the matrix
\begin{equation}
\label{GeometricSeries}
(\bI - {\bU} {\mathbf F})^{-1} - \bI \,\,\, = \,\,\,
  {\bU}{\mathbf F}
 +  ({\bU}{\mathbf F})^2
 +  ({\bU}{\mathbf F})^3
 + \cdots  +   ({\bU}{\mathbf F})^n,
\end{equation}
where $\bI$ is the $m \times m$ identity matrix.
We summarize our discussion in the following proposition,
which is proved by elementary matrix algebra.

\begin{prop}
\label{ZeroUnless}
The entry of the matrix (\ref{GeometricSeries})
in row $g$ and column $h$ is zero unless
$g \subset h$, in which case it is
$\,u_{gh} \cdot {\mathbf f}(h) \cdot P_{gh}({\mathbf f}) \,$
where $P_{gh}$ is a polynomial function of degree
$|h \backslash g|-1$ on the space
of all fitness landscapes $\,\mathbb{R}^\cG $.
\end{prop}

The polynomial  $\,P_{gh}({\mathbf f}) \,$  is the
generating function for all chains from $g$ to $h$ in $\cG$.
This will be made precise in the following corollary.
We shall restrict ourselves to the most important case
when  $ g = \hat{0}$ is the wild type
and $h = \hat{1}$ is the escape state.
Studying $\,P_{\hat{0} \hat{1}}({\mathbf f})\,$ only
is no loss of generality because any
interval of a distributive lattice
is again a distributive lattice.

Proposition~\ref{ZeroUnless} tells us
that $\,P_{\hat{0} \hat{1}}({\mathbf f})  \,$
is a polynomial of   degree $n-1$
in the unknown fitness values ${\bf f}(g)$,
which are also written as $f_g$, where $g \in \cG$.

\begin{cor} \label{AllThoseChains}
The polynomial $\,P_{\hat{0} \hat{1}}({\mathbf f})  \,$
in the upper-right entry of
(\ref{GeometricSeries}) equals
\begin{equation}
\label{RISK}
P_{\hat{0} \hat{1}}({\mathbf f}) 
\quad  = \sum_{\hat{0}=g_0 \subset g_1 \subset \dots \subset g_k = \hat{1}} 
\!\!\!\!\!\!  f_{g_1} f_{g_2} \cdots f_{g_{k-1}},
\end{equation}
where the sum runs over all chains
from $\hat{0}$ to $\hat{1}$ in 
the genotype lattice $\cG$.
\end{cor}


\section{The risk of escape}   \label{sec:branching}

For a poset of events $\cE$ and 
the corresponding distributive lattice $\cG = J(\cE)$, 
the \emph{risk polynomial} of $\cG$ is defined as the
polynomial (\ref{RISK}), which we denote by $\,\RP(\cG;{\mathbf f})$.
The risk polynomial was introduced
  in \cite{Iwasa2003,Iwasa2004}.
In this section we review the evolutionary dynamics model
proposed in these papers, and we
discuss the probabilistic meaning
of the risk polynomial. 

\begin{ex} \label{ex:rp} \rm
Let $\cG$ be the genotype lattice in Figure~\ref{fig:ex1}.
Then the risk polynomial
$\,\RP(\cG;{\mathbf f})\,$ is the following 
polynomial of degree three in six unknowns:
\begin{eqnarray*}
& 1 + f_{1000} + f_{0100} + f_{1100} + f_{0101} + f_{1110} + f_{1101} \\
&  + f_{1000} f_{1100} + f_{0100} f_{1100} + 
    f_{0100} f_{0101}  + f_{1000} f_{1110} + f_{0100} f_{1110}
 \\ &   + f_{1000} f_{1101} + f_{0100} f_{1101} +  f_{1100} f_{1110}
  + f_{1100} f_{1101} + f_{0101} f_{1101} \\
&   + f_{1000} f_{1100}  f_{1110} + f_{0100} f_{1100} f_{1110} 
 + f_{1000} f_{1100} f_{1101}  \\ &
     +  f_{0100} f_{1100} f_{1101} + f_{0100} f_{0101} f_{1101}.
\end{eqnarray*}
\end{ex}

If we restrict the fitness landscape
 ${\mathbf f}$ to lie in a linear subspace of $\mathbb{R}^\cG$,
then $\,\RP(\cG;{\mathbf f})$ specializes to a polynomial in fewer unknowns.
For example, the risk polynomial for graded fitness landscapes 
is obtained from the specialization
${\mathbf f}(g) = a_{|g|}$. That risk polynomial has
degree $n-1$ and is denoted by $\RP(\cG; a_1,\ldots,a_{n-1})$.
For instance, $\,\RP(\cG;{\mathbf f})\,$ in
Example \ref{ex:rp} specializes to
\[
   \RP(\cG; a_1,a_2,a_3) =
       1 + 2 a_1 + 2 a_2 + 2 a_3
     + 3 a_1 a_2 + 4 a_1 a_3 
     + 3 a_2 a_3 + 5 a_1 a_2 a_3.
\]
For constant fitness landscapes 
$\, {\mathbf f} \equiv a \,$, the risk polynomial is a polynomial in one unknown $\,a$.
It is denoted $\RP(\cG; a)$. In our running example,
\[ 
   \RP(\cG; a) \, = \, 1 + 6 a + 10 a^2 + 5 a^3. 
\]

We now make precise the notion of {\em risk of escape}, which will
justify our definition of the risk polynomial. 
Our derivation is based on the model 
for the dynamics of a replicating population
on a fitness landscape studied by
Iwasa, Michor and Nowak  \cite{Iwasa2003,Iwasa2004}.
See also the work of Wilke \cite{Wilke2003}
and the references given therein for approaches
to computing fixation probabilities.

A {\em multistate branching process} \cite{Athreya1972} consists
of a set of genotypes along with a fitness landscape and mutation
rates between genotypes.  We assume a discrete time process, where
in one generation an individual with genotype $g$
has a random number of offspring following a Poisson distribution
with mean $R_g$.  Some of these offspring may be mutants according to
the mutation rates $u_{gh}$.
The parameter $R_g$ is the {\em basic
reproductive ratio} \cite[Chap.~3]{Nowak2000}.

We assume there is no interaction between individuals; each reproduces
at a rate independent of the distribution of the population.
Let $\muta{g}{h}{k}$ be the probability
that one individual of genotype $g$ has $k$ children of type $h$.  Then, \begin{equation}\label{eq:1}
\muta{g}{h}{k}  \, = \, 
\frac {(u_{gh}R_g)^k \cdot e^{-u_{gh}R_g}} {k!}.
\end{equation}
The {\em reproductive fitness} $f_g$ is related to 
the reproductive ratio $R_g$ by
\begin{equation}
\label{Randf}
 f_g \, = \,  \frac {R_g} {1-R_g}
\qquad \hbox{and} \qquad
R_g \, = \, \frac{f_g}{1+f_g}. 
\end{equation}

Let $\xi_g$ be the probability of escape  starting with one individual of
genotype $g$, so $1 - \xi_g$ is the probability of extinction. 
In particular, $\xi_{\hat{1}}$ is the probability that one resistant
virus will not become extinct.
Each of these probabilities is a function
of the mutation rates $u_{gh}$ and the reproductive ratios $R_g$.
We assume that the $u_{gh}$ are as in 
(\ref{eq:muta}), but with $u_{gg} = 1$.
Thus, each escape probability $\xi_g$  can be expressed
as a function of the $\mu_e$
for $e \in \cE$ and  (using the relation (\ref{Randf}))
 the fitness values $f_g$ for $g \in \cG$.

\begin{thm} \label{thm:1}
If $\xi_g \ll 1$ for $g \neq \hat{1}$, then
the probability of escape on
the fitness landscape $\mathbf{f} \in \mathbb{R}^{\cG}$ starting with one
individual of wild type $\hat{0}$, satisfies
\begin{equation} \label{eq:2}
  \xi_{\hat{0}} \quad  \approx  \quad \xi_{\hat{1}} \cdot f_{\hat{0}} \cdot
    \prod_{e \in \cE} {\mu_e} \cdot \RP(\cG;\mathbf{f}).
\end{equation}
\end{thm}

\begin{proof}
The probability of extinction
satisfies the recursive formula
\begin{equation}
\label{michorproof}
   1-\xi_g \quad = \quad  \prod_{h \supseteq g} \sum_{k=0}^{\infty}
(1-\xi_h)^k \cdot 
       \muta{g}{h}{k} .
  \end{equation}
Using (\ref{eq:1}), the right hand side
of (\ref{michorproof}) can be rewritten as follows:
\begin{equation*}
\label{michorproof2}
      \prod_{h \supseteq g} 
       {\rm exp}({(1-\xi_h)u_{gh}R_g} ) \cdot {\rm exp} ({-u_{gh}R_g}) 
     \quad = \quad \exp\left(\sum_{h \supseteq g} -\xi_h u_{gh} R_g\right). 
\end{equation*}
We conclude that
\[
   \log(1-\xi_g) \quad = \quad - \sum_{h\supseteq g} \xi_h u_{gh} R_g \quad
\qquad \hbox{for all} \,\, g \in \cG.
\]
Under the assumption that $\xi_g \ll 1$ for $g \neq \hat{1}$, we can 
linearize the logarithms using
the relation $\,\log(1-\xi_g) \approx -\xi_g$. This implies,
for $\, g \in \cG \backslash \{\hat{1}\}$,
\begin{eqnarray*}
   \xi_g \quad  \approx  & R_g \cdot \sum_{h \supseteq g} \xi_h u_{gh} \\
  \quad    = & \frac {R_g} {1-R_g u_{gg}} \cdot \sum_{h \supset g} \xi_h u_{gh} \\
     = & f_g \cdot \sum_{h \supset g} \xi_h u_{gh}.
\end{eqnarray*}

The theorem now
follows by setting $g = \hat{0}$ and expanding the last equation recursively.
Here we are using the fact  from (\ref{eq:muta}) that the
product of the $u_{gh}$ over any
chain from $\hat{0}$ to $\hat{1}$ in $\cG$ 
equals $\,\prod_{e \in \cE} \mu_e$.
\end{proof}

The typical situation of interest is a fitness landscape for which
only the escape state has a basic reproductive ratio greater than one,
i.e.,
\[
   R_{\hat{1}} > 1 \qquad \mbox{and} \qquad 
   R_g < 1 \quad \mbox{for all} \quad g \not= \hat{1}.
\]
When the positive numbers $R_g$ are very small for 
 $g  \in \cG \backslash \{\hat{1}\}$ then the approximation 
(\ref{eq:2}) is valid, and
it shows the crucial role that the risk polynomial
$\RP(\cG;\mathbf{f})$ plays in 
assessing the risk of escape from the wild type $\hat{0}$
to the escape state $\hat{1}$.
The theorem implies that the risk of escape 
of a population of $N$ wild type viruses   
is $(1-\xi_{\hat{0}})^N$. In Section~\ref{sec:discussion} we
discuss the situation in which the population is not homogeneous
at the time of intervention.

\smallskip

The risk of escape is an important quantity in analyzing the
invasiveness of pathogens and in assessing
the success probability of medical interventions such as
chemotherapy. However, putting this concept into practice
depends on our ability to actually compute the risk polynomial. 
It turns out that methods from algebraic combinatorics lead
to efficient algorithms for this task.
In the Appendix, several methods are presented in detail.

\begin{figure}
\centering
\includegraphics[width=.6\textwidth]{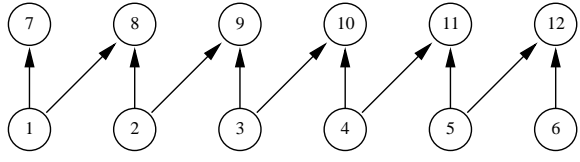}
\caption{Example of an event poset whose general risk polynomial 
is of degree 11 in 375 unknowns.}
\label{fig:poset}
\end{figure}

Our method of choice from a practical perspective
relies on computing linear extensions 
of the event poset $\cE$ (Theorem~\ref{thm:linearExtensions}, Appendix). 
Our software implementation is available at
\url{http://bio.math.berkeley.edu/riskpoly/} .  
For an example of the efficiency of the software, 
let $\cE$ be the poset in Figure~\ref{fig:poset} 
on $n=12$ events with cover relations
$i < 6 + i$ for $1 \leq i \leq 6$ and $i < 7 + i$ for $1 \leq i \leq 5$.
Here the genotype lattice $\cG$ consists of $375$ genotypes.
The risk polynomial $\RP(\cG; {\mathbf f})$ is a polynomial
of degree 11 in 375 unknowns $f_g$. 
This polynomial has 224,750,298 monomials in the 375
unknowns, but we represent it as a sum of
2,702,765 products, one for each 
linear extension of the event poset $\cE$.
Our software takes about ten seconds to compute
this representation of $\RP(\cG; {\mathbf f})$.
The result takes up 200MB of disk space.

 The univariate risk polynomial for this example is
\begin{multline*}
1 + 375a + 19088a^2 + 324498 a^3 + 2610169 a^4 + 11729394 a^5 +
32080336 a^6 +\\ 55597909 a^7 + 61448965 a^8 + 42020208 a^9 + 16216590
a^{10} + 2702765a^{11}.
\end{multline*}
Thus, exact symbolic computations, as opposed to numerical approximations,
may be necessary and feasible when one is interested in
assessing the risk of escape in applications like
the one described in Section \ref{sec:apply} below.


\section{Distributive lattices from Bayesian networks}   \label{sec:bayes}

In this section, we present a family of statistical models that naturally
gives rise to distributive lattices.  This statistical interpretation
provides a method for deriving the genotype lattice $\cG$ directly from data.
The basic idea is to estimate the poset structure on $\cE$ from
observed genotypes, by applying model selection techniques
to a range of Bayesian networks, and to
 define $\cG$ as the set of all genotypes with
non-zero probability in the model.

We first make precise the derivation of a genotype space
from a statistical model.
Let $\cE$ be an unordered set of $n$ genetic events. 
The events are labeled by $1,2,\ldots,n$. Subsets
of $\cE$ are identified with binary strings $g \in \{0,1\}^n$.
They are the possible genotypes.
We consider binary random variables $X_{\cE} = (X_1, \dots, X_n)$,
where $X_e = 1$ indicates the occurrence of event $e$. 
Let $\Delta$ denote the $(2^n-1)$-dimensional simplex
of probability distributions on $ \{0,1\}^n$. A {\em statistical
model} for $X_{\cE}$ is a map $\,p  \colon \Theta \to \Delta$,
where $\Theta$ is some parameter space.
The $g$-th coordinate of $p$, denoted
$p_g$,  is the
probability of genotype $g \in \{0,1\}^n$ under the model $p$.
The \emph{induced genotype space} of the model
$\,p \colon \Theta \to \Delta\,$ is the set
$\cG_p $ of all strings $\, g \in \{0,1\}^n \,$ such that
$p_g$ is not the zero function on $\Theta$.
We regard $\cG_p$ as a poset ordered by inclusion.

Now consider a directed acyclic graph on the set of events $\cE$. 
We will also call this graph $\cE$.
The {\em Bayesian network model}, or directed acyclic graphical model,
defined by $\cE$ is the family of joint distributions
that factor as
\begin{equation*}   \label{eqn:bayesnet}
   \Pr(X_1, \dots, X_n) 
     \quad = \quad \prod_{e \in \cE} \Pr(X_e \mid X_{\pa(e)}),
\end{equation*}
where $\pa(e)$ denotes the set of parents of $e$ in $\cE$.
Equivalently, a Bayesian network is specified by a  set of conditional independence
statements. Each node is independent of its ancestors given its parents.
See \cite{Lauritzen1996} for an introduction  to the relevant
statistical theory and \cite{Garcia2005} for an algebraic perspective.

The parameters for a Bayesian network are specified by
providing, for each event $e \in \cE$,
a $2^{|\pa(e)|} \times 2$
matrix $\theta^e$. The matrix entries are
\[   \theta^e_{g_{\pa(e)},g_e} 
     \quad = \quad \Pr\left( X_e = g_e \mid X_{\pa(e)} 
     = g_{\pa(e)} \right),  
\]
for  $ \, g_{\pa(e)} \in \{0,1\}^{\pa(e)},
\, g_e \in \{0,1\}$.
These conditional probabilities satisfy
\begin{equation}
\label{SumToOne}
\theta^e_{g_{\pa(e)},0} \geq 0\,,\,\,\,
\theta^e_{g_{\pa(e)},1} \geq 0 \,\, \quad \hbox{and} \,\,\quad
  \theta^e_{g_{\pa(e)},0}\, +\, \theta^e_{g_{\pa(e)},1} \,\,= \,\,1 . 
\end{equation}
    
Set $d = \sum_{e \in \cE} 2^{|\pa(e)|}$ and $\Theta = [0,1]^d$.
The points in the cube $\Theta$ are identified with $n$-tuples
of matrices $\,\theta = (\theta^e \,|\, e \in \cE)\,$ as above.
The {\em general Bayesian network} is the polynomial map
$\, p  \colon  \Theta \, \rightarrow \,\Delta \,$
 whose coordinates are
  \begin{equation}   \label{eqn:bayesfactor}
   p_g(\theta) \,\,\,= \,\,\, \prod_{e \in \cE} \theta^e_{g_{\pa(e)}, g_e}.
\end{equation}
The general Bayesian network on $\cE$ induces the
genotype space $\cG_p = \{0,1\}^n$, the Boolean lattice on $\cE$.
Indeed, the factorization~(\ref{eqn:bayesfactor}) implies
that no genotype $g \in \{0,1\}^n$ has probability zero for all
parameter values. 

To obtain other genotype spaces, we replace the
cube $\Theta = [0,1]^d$ by one of its faces, as follows.
For each event $e \in \cE$ consider a Boolean function
$\,\beta_e \colon \{0,1\}^{\pa(e)} \rightarrow \{0,1\}$.
If $\beta_e(g_e) = 0$ then
the row of the $2^{|\pa(e)|} \times 2$-matrix $\theta^e$ 
indexed by the genotype $g$ is fixed
to be the vector $(1,0)$;
otherwise that row remains indeterminate
subject to the constraints (\ref{SumToOne}).
Let $\Theta^\beta$ denote the face of $\Theta$
determined by these requirements
and  $\,p^\beta \colon \Theta^\beta \,\rightarrow \,\Delta\,$
 the restriction of the polynomial map $p$ to $\Theta^\beta$.
The resulting model is the Bayesian network on $\cE$ constrained by the
Boolean functions $\beta^e$.

If all Boolean functions $\beta^e$ are disjunctions
then we get the {\em disjunctive Bayesian network} on $ \cE$.
In this model, an event $e$ can only occur if at least one
of its parent events has already occurred.
If all Boolean functions $\beta^e$ are conjunctions
then we get the {\em conjunctive Bayesian network} on $\cE$.
In this model, an event $e$ can only occur if all
of its parent events have already occurred.
These restricted Bayesian network models induce 
 interesting genotype spaces. 
Our main result in this section concerns the conjunctive case.

We regard the given directed acyclic graph $\cE$ as a poset by setting $e_1
\leq e_2$ if there exists a path from $e_1$ to $e_2$.
We write $\,p^{\rm conj} \colon [0,1]^n \rightarrow \Delta\,$
for the conjunctive Bayesian network on $\cE$,
since it has precisely $n$ free parameters.

\begin{thm} \label{fromBNtoDL}
The genotype space induced by the conjunctive
Bayesian network on $\cE$ is the distributive lattice of order ideals
in $\cE$, i.e., $\cG_{p^{\rm conj}} = J(\cE)$.
\end{thm}

\begin{proof}
The possible genotypes $g $ are binary strings whose coordinates $g_e$
indicate whether or not the event $e$ has occurred. If $p$ is
any of the Bayesian network models discussed above, then
 (\ref{eqn:bayesfactor}) implies that $g \in \cG_p$ if and only if
each $\thet{g}$ is non-zero. Consider now the
 conjunctive model $\,p = p^{\rm conj}$.
Here, the conditional probability
 $\thet{g}$ is non-zero if and
only if $g_e = 1$ implies $g_{\pa(e)} = (1, \dots, 1)$.  This is
precisely the condition for $g$ to be an order ideal in $\cE$.
Thus $\cG_p$ is the distributive lattice of order ideals of $\cE$.
\end{proof}

The following example illustrates Theorem \ref{fromBNtoDL},
and it compares the genotype spaces induced by 
the disjunctive and the conjunctive  Bayesian network.
The former is not a distributive lattice, 
but the latter always is.

\begin{ex} \label{ex:conjunctive} \rm
Let $\cE$ be the event poset in Figure~\ref{fig:ex1}.
The general Bayesian network model defined by $\cE$
is parametrized by the following four matrices:

\vspace{1ex}
\parbox{3.5cm}{ \centering
$
  \begin{array}{l}
    \theta^1 = 
    \left( \begin{array}{cc}
      a & 1-a 
    \end{array} \right), \\[2ex] 
    \theta^2 =
     \left( \begin{array}{cc}
      b & 1-b 
    \end{array} \right),
  \end{array}
$
}
\parbox{4.5cm}{ \centering
$
  \theta^3 = \left(
  \begin{array}{cc} 
   c_{00} & 1 - c_{00} \\
   c_{01} & 1 - c_{01} \\
   c_{10} & 1 - c_{10} \\
   c_{11} & 1 - c_{11}
  \end{array} \right),
$
}
\parbox{4cm}{ \centering
$
    \theta^4 =
     \left( \begin{array}{cc}
      d_0 & 1 - d_0 \\
      d_1 & 1 - d_1 
    \end{array} \right).   
$
}\\
\vspace{1ex}

\noindent The map $p \colon [0,1]^8 \to \Delta$ has coordinates
\begin{eqnarray*} &
 p_{0000} \, = \, a b c_{00} d_0, &
 p_{0001} \, = \, a b c_{00} (1-d_0) , \\ &
 p_{0010} \, = \, a b (1-c_{00}) d_0, &
 p_{0011} \, = \, a b (1-c_{00}) (1-d_0), \\ &
 p_{0100} \, = \, a (1-b) c_{01} d_1, &
 p_{0101} \, = \, a (1-b) c_{01} (1-d_1), \\ &
 p_{0110} \, = \, a (1-b) (1-c_{01}) d_1, &
 p_{0111} \, = \, a (1-b) (1-c_{01}) (1-d_1), \\ &
 p_{1000} \, = \, (1-a) b c_{10} d_0, &
 p_{1001} \, = \, (1-a) b c_{10} (1-d_0), \\ &
 p_{1010} \, = \, (1-a) b (1-c_{10}) d_0, &
 p_{1011} \, = \, (1-a) b (1-c_{10}) (1-d_0), \\ &
 p_{1100} \, = \, (1-a) (1-b) c_{11} d_1, &
 p_{1101} \, = \, (1-a) (1-b) c_{11} (1-d_1), \\ &
 p_{1110} \, = \, (1-a) (1-b) (1-c_{11}) d_1, &
 p_{1111} \, = \, (1\!-\!a) (1\!-\!b) (1 \! - \! c_{11}) (1 \! - \! d_1).
\end{eqnarray*}
This model induces the Boolean lattice $\{0,1\}^4$ as genotype space.

The disjunctive Bayesian network is the 
six-dimensional  submodel  obtained by setting
$\,c_{00}=1 \,$ and $ \,d_0=1 $. This substitution implies
\[
p_{0001} \,=\, p_{0010} \, = \, p_{0011} \,=\,
p_{1001}  \,=\, p_{1011} \,\, = \,\, 0.
\]
The genotype space $\,\cG_{p^{\rm disj}}$
consists of the remaining eleven strings in
$\{0,1\}^4$. Note that 
$\,\cG_{p^{\rm disj}} \,$ is not
a  lattice because it is not
closed under intersections. For instance,
$\,1010$ and $ 0110 $ are in $ \cG_{p^{\rm disj}} \,$ 
but $\,0010 =  1010\,\cap \, 0110
 \not\in \cG_{p^{\rm disj}} $.

The conjunctive Bayesian network is the 
four-dimensional  submodel  obtained by setting
$\, c_{00}= c_{01}= c_{10}= d_0 = 1$. The
remaining eight non-zero probabilities are
indexed by the eight genotypes in Figure~\ref{fig:ex1}:
\begin{eqnarray*}
&  p_{0000} \, = \, a b   \, ,\,\,&
 p_{0100} \, = \, a (1-b)  d_1 \, ,\,\,\\
& p_{0101} \, = \, a (1-b) (1-d_1) \, ,\,\, &
 p_{1000} \, = \, (1-a) b \,,\,\,\, \\
& p_{1100} \, = \, (1-a) (1-b) c_{11} d_1 \, ,\,\, &
 p_{1101} \, = \, (1-a) (1-b) c_{11} (1-d_1) \, ,\,\, \\
& p_{1110} \, = \, (1-a) (1-b) (1-c_{11}) d_1 \, ,\,\, &
 p_{1111} \, = \, (1\! - \! a) (1\! - \! b)
 (1 \! - \! c_{11}) (1 \! - \! d_1).
\end{eqnarray*}
\end{ex}

If $\cE$ is a directed forest, i.e.,
if every $e \in \cE$ has at most one parent,
then we can augment $\cE$ to a tree $\cE^T$
by adding an auxiliary root node $0$
which points to the roots (edges with no parents) of the forest.
On the resulting tree $\cE^T$ we consider the
{\em mutagenetic tree model} of \cite{Beerenwinkel2005f, Desper1999}.

\begin{prop}   \label{prop:forest}
If $\cE$ is a directed forest then the following three statistical
models coincide: the disjunctive Bayesian network on $\cE$,
the conjunctive Bayesian network on $\cE$, and the
 mutagenetic tree model on $\cE^T$.
\end{prop}

\begin{proof}
The disjunctive and the conjunctive networks 
coincide because they are defined by the same
specializations of the parameters $\,\theta^e$.
The identification with the mutagenetic tree model follows from
\cite[Thm.~14.6]{Beerenwinkel2005c}.
\end{proof}

Mutagenetic tree models can be learned from observed data by an efficient
combinatorial algorithm.
With appropriate edge weights that depend on the pairwise
probabilities of events, a mutagenetic tree can be obtained as the maximum
weight branching rooted at 0 in the complete graph on $\{0,\dots,n\}$; see
\cite{Desper1999}. This gives an efficient method for learning
the poset $\cE$, and hence the genotype lattice $\cG = J(\cE)$, from
data. It would be interesting to extend this model selection
technique to arbitrary  conjunctive Bayesian networks.



\section{Applications to HIV drug resistance}   \label{sec:apply}

We investigate the development of resistance during treatment of HIV
infected patients with two different PIs. Consider the seven genetic events
\[
   \cE \, = \, \left\{ \mbox{K20R,~M36I,~M46I,~I54V,~A71V,~V82A,~I84V} \right\},
\]
where K20R stands for the amino acid change from lysine (K) to arginine (R) 
at position 20 of the protease chain, etc. 
The occurrence of these mutations confers broad cross-resistance to the 
entire class of PIs. Appearance of the virus with
all 7 mutations renders most of the PIs ineffective for subsequent 
treatment.  We analyze the risk of reaching this escape state under 
therapy with the PIs ritonavir (RTV) and indinavir (IDV)
\cite{Condra1996, Molla1996}.

\begin{figure}[!tpb]
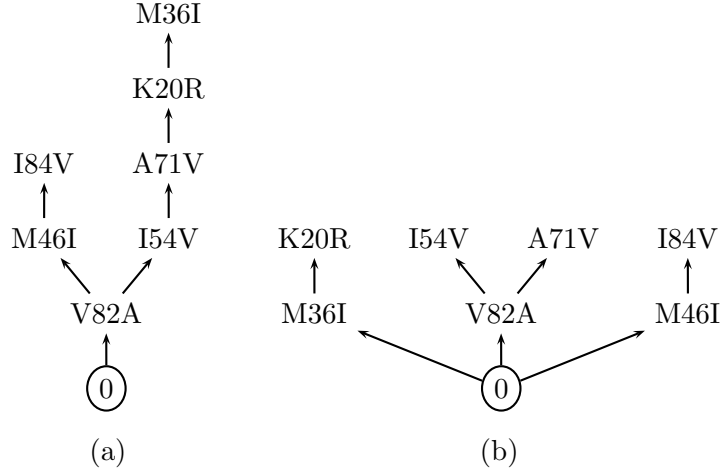

\centering
\begin{tabular}{ccc}
\tree{\root{0}}{
	\tree{\node{V82A}}{
		\tree{\node{M46I}}{
			\node{I84V}
			}
		\tree{\node{I54V}}{
			\tree{\node{A71V}}{
				\tree{\node{K20R}}{
					\node{M36I}
					}
				}
			}
		}
	}
& ~~~~~~~~~~~~~ &
\tree{\root{0}}{
	\tree{\node{M36I}}{
		\node{K20R}
		}
	\tree{\node{V82A}}{
		\node{I54V}
		\node{A71V}
		}
	\tree{\node{M46I}}{
		\node{I84V}
		}
	}\\[3ex]
(a) & & (b)
\end{tabular}
\caption{Mutagenetic tree $\cE^T$ for the development of resistance 
to (a) ritonavir and (b) indinavir in the HIV-1 protease.
The event poset $\cE$ is obtained by removing the 
root node ``0''.}
\label{fig:trees} 
\end{figure}

We use mutagenetic trees for estimating preferred mutational pathways
and for defining genotype lattices.
For both drugs, a tree $\cE^T$ is learned from genotypes derived 
from patients under the respective therapy. We used 112 and 691 samples
from the Stanford HIV Drug Resistance Database \cite{Rhee2003} 
for ritonavir and indinavir, respectively. 
Figure~\ref{fig:trees} shows the inferred mutagenetic trees.
The models indicate that the evolution of ritonavir
resistance is partly a linear process, whereas indinavir resistance
develops in a less ordered fashion. This is consistent with
previous studies \cite{Condra1996, Molla1996}.
The genotype lattices $\cG$  have size
$16 $  for ritonavir and $45$ for indinavir.
We study the risk polynomials on these 
lattices under different fitness landscape models.  

For the constant fitness landscape on $\,\cG \backslash 
\{\hat{0}, \hat{1}\}$, we obtain
\begin{eqnarray*}
  \RP_{\rm RTV}(a) &=& 15a^6+70a^5+131a^4+124a^3+61a^2+14a+1, \\
  \RP_{\rm IDV}(a) &=& 420a^6+1470a^5+1970a^4+1250a^3+372a^2+43a+1.
\end{eqnarray*}
Thus, the risk of developing all seven PI resistance mutations 
is higher under indinavir therapy than under ritonavir:
$  \RP_{\rm IDV}(a) >   \RP_{\rm RTV}(a)$ for $a > 0$.
Intuitively, the risk under ritonavir is lower because
the mutations must occur in a certain order. Likewise,
the high risk under indinavir results from many mutations occurring
independently, which gives rise to a large genotype lattice and to many
mutational pathways from the wild type to the escape state. 

More realistic fitness landscapes may be derived by modeling viral fitness
as a function of drug concentration. We follow the approach pursued
in \cite{Stilianakis1997a} and use a simple saturation function for
this dependency. Specifically, we assume viral fitness to be the following
function of drug concentration $D$,  
\begin{equation}   \label{eqn:drugfitness}
   f_g(D) \quad = \quad \frac{\phi_g}{1 + D/r_g},
\end{equation}
where $\phi_g$ denotes the fitness of genotype $g$ in the absence of drug
and $r_g$ the IC$_{50}$ value of $g$, i.e., the drug concentration necessary
to inhibit viral replication \emph{in vitro} by 50\%. The IC$_{50}$ value 
is a measure of resistance. We will assume
throughout that all $\phi_g \equiv \phi$ are equal.
If we assume, in addition,
that the resistance landscape is constant on $\cG \backslash \{\hat{0},\hat{1}\}$,
with $r_g \equiv r$,
then the substitution (\ref{eqn:drugfitness}) turns
the risk polynomial into a rational function in $\phi$, $D$, and $r$.
For example, for ritonavir, this rational function is
\[
   \frac{(15\phi^2r^2+10\phi Dr+10\phi r^2+D^2+2Dr+r^2)(\phi r+D+r)^4}{(D+r)^6}.
\]

\begin{figure}
\centering
\includegraphics[width=.8\textwidth,angle=270]{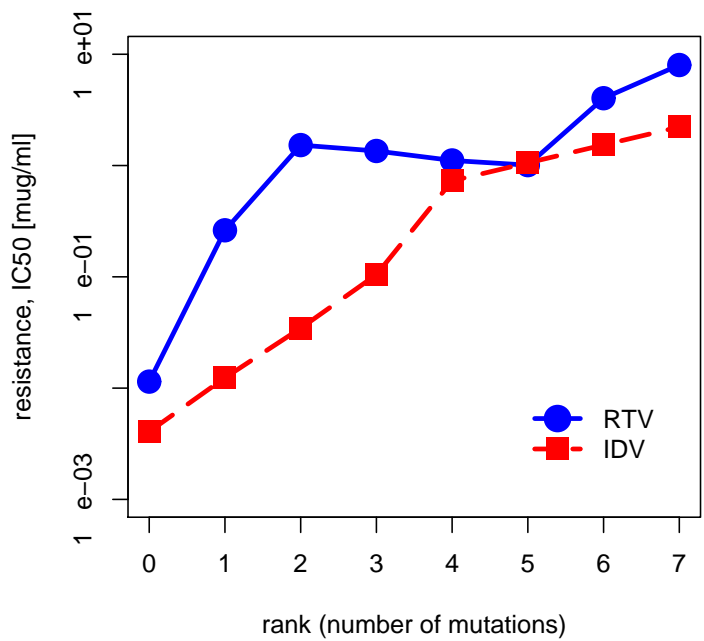}
\caption{Graded resistance landscapes for ritonavir (RTV, bullets)
and indinavir (IDV, squares). Resistance is quantified as the
drug concentration necessary to inhibit viral replication \emph{in vitro}
by 50\% (IC$_{50}$).}
\label{fig:gfl}
\end{figure}

In general, the IC$_{50}$ values $r_g$ are distinct and can be determined
experimentally for some genotypes
by phenotypic resistance testing \cite{Walter1999},
and may be predicted for all genotypes using regression techniques
\cite{Beerenwinkel2003d}.
PI phenotypic resistance data suggests a graded resistance landscape;
see \cite{Berkhout1999} and \cite[Tab.~3]{Condra1996}.
Hence, we estimate the resistance $r \in \mathbb{R}^8$
for ritonavir and indinavir by defining $r_k$ 
as the mean predicted IC$_{50}$ of all 
genotypes of rank~$k$. The resulting resistance landscapes
are shown in Figure~\ref{fig:gfl}.
  
\begin{figure}
\centering
\includegraphics[height=\textwidth,angle=270]{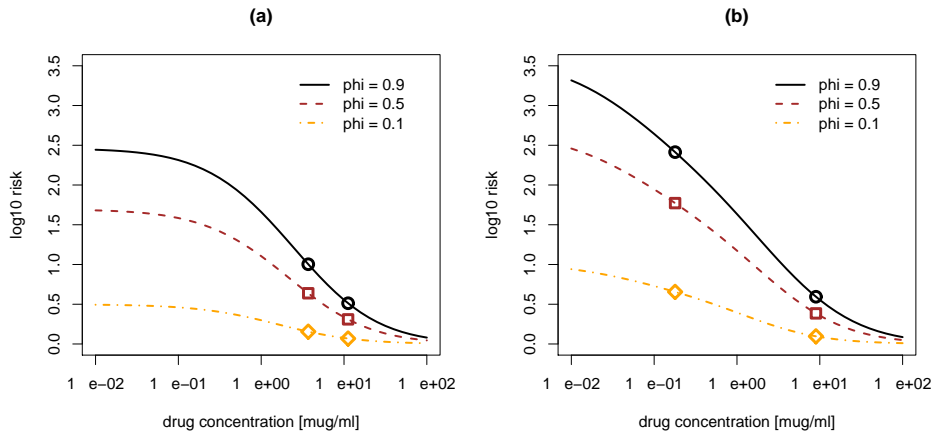}
\caption{Drug dependent risk. The log of the risk polynomial
for ritonavir (a) and indinavir (b) 
is displayed as a function of plasma drug concentration $D$. Marked
values denote mean trough ($C_{\min}$) and peak ($C_{\max}$)
levels observed in clinical studies. The parameter $\phi$ is
the relative fitness of mutants as compared to the wild type
in the absence of drug.}
\label{fig:drugfit}
\end{figure}

The graded risk polynomials $\RP(a_1,a_2,a_3,a_4,a_5,a_6)$ have 64 terms. After
substituting $a_k = \phi/(1 + D/r_k)$, we obtain rational risk functions in $D$
with parameter $\phi$. Figure~\ref{fig:drugfit} illustrates the dependency of
the risk on drug concentration for three different values of $\phi$. For both
drugs we indicate published mean plasma trough ($C_{\min}$) and peak ($C_{\max}$) levels 
observed in clinical settings.

This example illustrates how the risk
polynomial can be used to study viral escape as a function of
different parameters. For instance, given a pharmacokinetics model
of antiretroviral drug therapy, we can compute
the risk of developing resistance after a patient has missed a dose. 
Thus, our mathematical framework may help in designing robust drug combinations.


\section{Discussion}   \label{sec:discussion}

We have presented a computational framework for assessing
the risk of escape of an evolving population of pathogens.
The risk of escape is the probability that the population
reaches an escape state before extinction.
In virus transmissions, for example, this probability is 
the chance of survival in the new host. In the situation
of antiretroviral therapy, the risk of escape is the
probability of therapy failure due to the development
of drug resistance.

The general setup we consider for computing the risk of escape
includes an event poset, a fitness landscape on its induced
genotype lattice, and a branching process on this lattice.
The event poset $\cE$ consists of all mutational events that can
occur and encodes the constraints which apply to their order of 
occurrence. From this structure the genotype space $\cG$ is obtained
by considering all mutational pathways that respect the order
constraints. This natural construction endows $\cG$ with the
mathematical structure of a distributive lattice.
The risk polynomial, the crucial factor in
computing the risk of escape, turns out to coincide with the chain
polynomial of the genotype lattice. We have presented
methods from algebraic combinatorics that exploit
this connection and that result in efficient algorithms.     

The space of genotypes may also be inferred from 
observed genotype data using statistical model selection tools.
We have identified a class of Bayesian network models,
the conjunctive Bayesian networks, whose support induces
a genotype lattice. 
Mutagenetic tree models arise as important special cases.
Here, both statistical model selection
and risk computation are particularly efficient, and readily available
with existing software \cite{Beerenwinkel2005b} 
coupled with our implementation of the linear extensions
method (Theorem~\ref{thm:linearExtensions}, Appendix). 

\smallskip

We have focused on the dependency of the risk polynomial
on the fitness landscape and considered throughout a homogeneous
wild type population prior to intervention. However, the risk of
escape is calculated  similarly for a quasispecies 
distribution at the time of intervention. In fact,
this involves computing the risk polynomial of
the prior fitness landscape \cite{Iwasa2003}.
In contrast, the branching process
model can not account  
for recombination, horizontal gene transfer, or frequency 
dependent selection, since evolution is assumed to take place
in multiple lineages independently. 

The main challenge in using our method to compute the risk
of escape from antiretroviral therapy lies in accurately
modeling the fitness landscape. 
The dependency (\ref{eqn:drugfitness}) of the fitness on drug 
concentration may be improved by experimentally determined
viral replicative capacities in the
absence of drugs. An alternative approach to derive a 
fitness landscape for HIV-1 proteases is based on estimating
the binding affinity of the drug to the mutant protease, and
the mutant's ability to cleave its natural substrates
\cite{Rosin1999a}. 
These calculations are based on simplified molecular
modeling techniques. 
The resulting fitness landscape does not account for different
drug levels, but it is independent of experimental 
resistance and fitness data.
 
Escape from indinavir and ritonavir therapy may in some cases
involve mutations other than the seven we considered, although those
are the most frequent mutations observed after therapy failure
\cite{Condra1996,Molla1996}.
On the other hand, viral escape might be accomplished with
genotypes that harbor fewer than all of the mutations.
Thus it would be desirable to compute the risk of reaching
any of several escape states, rather than only the $11\cdots 1$ type.
This computation will involve similar techniques to those presented
in Section~\ref{sec:branching} and the Appendix.

Finally, the PIs form only one out of four distinct
classes of antiretroviral drugs
that are in current clinical use. The standard of care is combination
therapy with at least three different drugs from two different drug 
classes. Modeling the fitness landscape of combination therapy in
terms of viral drug resistance and drug exposure is even more
challenging, but can eventually help in designing optimal 
antiretroviral therapies.  Algebraic combinatorics offers
tools for the mathematical analysis of these 
biomedical problems.


\section*{Acknowledgements} 

Niko Beerenwinkel is supported by Deutsche Forschungsgemeinschaft under
grant No.\ BE~3217/1-1.
Nicholas Eriksson and Bernd Sturmfels are supported by
the U.S.~National Science Foundation,
under the grants  EF-0331494 and DMS-0456960
respectively, and by the DARPA program
{\em Fundamental Laws in Biology} (HR0011-05-1-0057).


\bigskip

\bibliographystyle{plain}

\section*{Appendix: Mathematics and computation of the risk polynomial}

Here we discuss in more detail mathematical properties 
of the risk polynomial and we present several methods for computing it.
The given data consists of an $n$ element poset $\cE$ 
and its induced genotype lattice $\cG$, which is the distributive 
lattice of order ideals in $\cE$. We assume that $\cG$ has 
$m$ elements, which are encoded either
as subsets of $\cE$ or as binary strings in $\{0,1\}^n$.
The risk polynomial is the polynomial $\,\RP(\cG;{\bf f})\,$
in the $m$ unknowns $f_g = {\bf f}(g)$, 
one for each genotype $g$.
We are also interested in  specializations of
$\RP(\cG;{\bf f})$ obtained by setting some (or all) of the unknowns
equal to each other, such as
the graded risk polynomial and the univariate risk polynomial.

\subsection*{Stanley's linear algebra method}

A direct method for computing the risk polynomial is given 
in Section~\ref{sec:branching}.
 Namely, we can set all $\mu_e$  equal to one
in the matrix ${\bf U}$ and then compute the upper right
entry of the matrix $\,({\bf I} - {\bf UF})^{-1} - {\bf I} \,$ of
equation (\ref{GeometricSeries}).
In practice, one would compute this entry
by a dynamic program which runs in time $O(m^2)$.
That dynamic program is easily   derived by resolving the recursion
in  the last equation of the proof  of Theorem~\ref{thm:1}.

The following alternative linear algebra technique for 
computing polynomials similar to our risk polynomials 
was given by Stanley  in \cite{Stanley1996}.
Let $\,\cG' = \cG  \backslash \{\hat{0}, \hat{1}\} \,$ denote
the genotype lattice with the top element
$\hat{1}$ and the bottom element $\hat{0}$ removed.
We define ${\bf A} $ to be the {\em anti-adjacency matrix} of the truncated 
genotype lattice $\cG'$. Thus ${\bf A}$ is the $(m-2) \times (m-2)$-matrix
with rows and columns indexed by $\cG'$, and whose entry
in row $g$ and column $h$ is $0$ if $ g \subset h$
and is $1$ otherwise. We write ${\bf I}$ for the 
$(m-2) \times (m-2)$ identity matrix and 
$\, {\bf F}' = {\rm diag} \bigl( \,{\bf f}(g) \,|\, g \in \cG' \bigr)\,$ for the
$ (m-2) \times (m-2)$-diagonal matrix whose entries are the
fitness values. Stanley's result reads as follows.

\begin{thm}[Stanley \cite{Stanley1996}]
\label{stanley}
The risk polynomial $\,\RP(\cG; {\bf f})\,$ equals
the determinant of the $(m-2) \times (m-2)$-matrix $\, {\bf I} \, +\, {\bf F}' \cdot {\bf A}$.
\end{thm}

\begin{ex} \rm
Let $\cG$ be the genotype lattice in Figure~\ref{fig:ex1}. Then $m =8$ and
 $\, {\bf I} \, +\, {\bf F}' \cdot {\bf A}\,$ is the $6 \times 6$-matrix
\[
\bordermatrix{ &  1000 & 0100 & 1100 & 0101 & 1110 & 1101 \cr
1000 & 1 + f_{1000} &   f_{1000} &    0  &   f_{1000} &   0 &   0 \cr
0100 &       f_{0100} &    1 + f_{0100} &    0  &    0  &    0 &    0 \cr
1100 & f_{1100} &   f_{1100} &   1 + f_{1100 } &   f_{1100 } & 0 &  0 \cr
0101 &  f_{0101} & f_{0101} &  f_{0101 } & 1 + f_{0101} &  f_{0101} & 0 \cr
1110 &  f_{1110 } &  f_{1110} &  f_{1110} & f_{1110} &  1 + f_{1110} &  f_{1110} \cr
1101 & f_{1101} &  f_{1101} & f_{1101} &  f_{1101} &  f_{1101} &  1 + f_{1101} \cr}.
\]
The determinant of this matrix is
the risk polynomial of Example~\ref{ex:rp}.
\end{ex}

\subsection*{The Hilbert series method}

A more conceptual way of thinking about the risk polynomial
is based on the following algebraic construction.
The {\em Stanley-Reisner ideal} $\,I_{\cG'}\,$ of $\cG'$
is the ideal generated by all quadratic monomials
$\,f_g \cdot f_h \,$ where $g$ and $h$
are genotypes that are incomparable,
i.e., neither $g \subseteq h$ nor $h \subseteq g$ holds.
The ambient polynomial ring $\,S = \mathbb{R}[{\bf f}] $ 
is generated by the unknowns $f_g$ where $g \in \cG'$. 
The {\em Hilbert series} of $\,I_{\cG'}\,$
is the formal sum over all monomials
$\,{\bf f}^u \, = \,\prod_{g \in \cG'} f_g^{u_g}\,$
which  are not in the ideal $\,I_{\cG'}$.
This is a formal generating function which can be
written as a rational function of the following form
\[
H(S/I_{\cG'}; {\bf f}) \quad = \quad
\frac{K_\cG({\bf f})}{\prod_{g \in \cG'} (1-f_g)}.
\]
Here $K_\cG({\bf f})$ is a polynomial
in the unknowns $f_g$ with integer coefficients.
The polynomial $K_\cG({\bf f})$
is known as the {\em K-polynomial} of the ideal $I_{\cG'}$.
We refer to \cite{Miller2004} for an introduction
to Stanley-Reisner ideals and their K-polynomials.

If $\cE$ is a directed forest (and we identify $f_g = p_g$) 
then Proposition \ref{prop:forest} and
\cite[Thm.~14.11]{Beerenwinkel2005c} imply that
the ideal $I_{\cG'}$ is an initial monomial ideal
of the conjunctive Bayesian network on $\cE$.
In a forthcoming paper we shall prove
that this initial ideal property holds
for all event posets (not just trees).

\begin{ex} \rm
Let $\cG$ be the genotype lattice in Figure~\ref{fig:ex1}.
Then 
\[
I _{\cG'} \quad = \quad \langle\,
f_{0101} f_{1110},\,
f_{1101} f_{1110},\,
f_{0101} f_{1100},\,
f_{0101} f_{1000},\,
f_{0100} f_{1000}
\rangle
\]
is indeed the initial monomial ideal
of the conjunctive Bayesian network
in Example~\ref{ex:conjunctive}.
The K-polynomial $K_{\cG}({\bf f})$ equals
\begin{eqnarray*}
& 1
- f_{0101} f_{1110}
- f_{1101} f_{1110}  
- f_{0101} f_{1100}
- f_{0101} f_{1000}
- f_{0100} f_{1000} \\ &
+ f_{0100} f_{1000} f_{0101}
+ f_{1000} f_{0101} f_{1100}
+ f_{1000} f_{0101} f_{1110}
+ f_{0101} f_{1100} f_{1110} \\ &
+ f_{0101} f_{1110} f_{1101}
+ f_{0100} f_{1000} f_{1110} f_{1101} \\ &
- f_{1000} f_{0101} f_{1100} f_{1110}
- f_{0100} f_{1000} f_{0101} f_{1110} f_{1101}.
\end{eqnarray*}
\end{ex}

\smallskip

Again using Proposition~\ref{prop:forest} and 
Theorem~14.11 in \cite{Beerenwinkel2005c}
we see that 
the risk polynomial  $\,\RP(\cG; {\bf f})\,$
is the sum of all squarefree monomials
in the expansion of the Hilbert series $H(S/I_{\cG'}; {\bf f})$.
Equivalently, $\,\RP(\cG; {\bf f})\,$ is the reduction of
$H(S/I_{\cG'}; {\bf f})$ modulo the ideal generated
by the squares $\,f_g^2 \,$ of the unknowns.
Since $\,1/(1-f_g)\,$ equals $\,1+f_g\,$ modulo 
$\,\langle \, f_g^2 \, \rangle $, we have the following result.

\begin{prop} \label{reisner}
The risk polynomial  $\,\RP(\cG; {\bf f})\,$ 
of the genotype lattice $\cG$ is the sum of
all squarefree terms in the expansion of
\[
K_\cG({\bf f}) \cdot \prod_{g \in \cG'} (1+f_g),
\]
where $K_\cG({\bf f})$ is the $K$-polynomial
of the Stanley-Reisner ideal $I_{\cG'}$.
\end{prop}

The univariate risk polynomial $\,\RP(\cG; a) \,$
 is derived from $\,\RP(\cG;{\bf f})\,$
by replacing each $f_g$ by the scalar unknown $a$.
We have
\[
\RP(\cG;a) \quad = \quad
c_0 + c_1 a + c_2 a^2 + \cdots + c_{n-1} a^{n-1},
\]
where $c_i$ is the number of chains of length $i$ in $\cG'$. Thus, 
$(c_0,\ldots,c_{n-1})$ is the $f$-vector
of the simplicial complex of chains in $\cG'$.
Likewise, we get the graded risk polynomial from
$\RP(\cG;{\bf f})$ by replacing each $f_g$ by
$a_{|g|}$. We note that the graded risk polynomial is  related to
Ehrenborg's quasi-symmetric function encoding \cite{ehrenborg1996}
of the flag $f$-vector of the chain complex of $\cG'$.

\subsection*{The linear extensions method}

One advantage of both Theorem~\ref{stanley}
and Proposition~\ref{reisner} is that these
formulas do not actually depend on the
fact that $\cG$ is a distributive lattice.
They also apply if the set
$\cG$ of genotypes is an arbitrary
poset. This is relevant for our
discussion of the statistical models in Section~\ref{sec:bayes},
where we introduced a more general
class of posets $\cG_p \subseteq \{0,1\}^n$.

This advantage is also a disadvantage: 
Theorem~\ref{stanley} and Proposition~\ref{reisner}
do not give the most efficient methods for
computing  $\RP(\cG;{\bf f})$ when $\cG$ is  the distributive lattice
induced by an event poset $\cE$. In what follows
we present a specialized and more efficient
algorithm for the risk polynomial.
 The input to this algorithm consists of
the event poset $\cE$. It is not necessary
to compute the genotype lattice $\cG$ 
as this will be done as a byproduct of our approach,
which is to compute  the risk polynomial $\RP(\cG;{\bf f}) $ directly from $\cE$.
  
As before, we assume that $\cE$ has $n$ elements, and 
we write $[n]$ for the linearly ordered set $\{1,2,\ldots,n\}$.
A {\em linear extension} of $\cE$ is an order-preserving
bijection $\,\pi \colon \cE \rightarrow [n]$. This means that
$e < e'$ in $\cE$ implies $\pi(e) < \pi(e')$.
Every linear extension  $\,\pi \colon \cE \rightarrow [n]$
gives rise to an ordered list of $n-1$ genotypes 
$\,g^{(1)},g^{(2)}, \ldots,g^{(n-1)}\,$ in
$\,\cG' = \cG \backslash \{\hat{0},\hat{1}\}$ as follows.
The genotype $g^{(i)}$ is
the subset of $\cE$ consisting of all
events whose image under $\pi$ 
is among the first $i$ positive integers. In symbols,
$\, g^{(i)} \,= \, \pi^{-1}(\{1,2,\ldots,i\}) $.
The sequence $g^{(1)}, g^{(2)}, \ldots, g^{(n-1)}$, derived from $\pi$,
represents a mutational pathway in $\cG$.

We now fix one distinguished linear extension of $\cE$,
that is, we identify the set underlying $\cE$ with $[n]$ itself.
Then a linear extension is simply
any permutation $\pi$ of $[n]$ which preserves the
order relations in $\cE$.  We define
\begin{equation}
\label{FPi}
{\bf f}(\pi) \quad = \quad 
\prod_{i: \pi(i) < \pi(i+1)} ( f_{g^{(i)}} + 1) 
\cdot
\prod_{i: \pi(i) > \pi(i+1)}  f_{g^{(i)}}  ,
\end{equation}
where $i$ runs over $\{1,2,\ldots,n-1\}$.
Our algorithm amounts to evaluating
the  risk polynomial by means of the
following explicit summation formula.

\begin{thm}   \label{thm:linearExtensions}
The risk polynomial  $\RP(\cG;{\bf f}) $ 
equals the sum of the products ${\bf f}(\pi)$
where $\pi$ runs over all linear extensions of 
the event poset $\cE$.
\end{thm}

\begin{proof}
The relationship between chains in $\cG$ and
linear extensions of $\cE$ is the content of
\cite[Prop.~3.5.2]{Stanley1999}.
The distributive lattice $\cG$ has a canonical
{\em R-labeling} \cite[Sec.~3.13]{Stanley1999}
which assigns to each edge of the Hasse diagram of
$\cG$ the corresponding element of $\cE$.
In view of this R-labeling, Exercise~59d in \cite[Chap.~3]{Stanley1999} 
tells us that the poset $\,\cG'  = \cG \backslash \{\hat{0},\hat{1}\}\,$ is
{\em chain-partitionable}.
Each product  ${\bf f}(\pi)$ as in (\ref{FPi})
 is the generating function for
 all the chains in precisely one part of that chain
partition of $\cG'$. Adding up all products
gives the generating function for all chains,
which is the risk polynomial.
\end{proof}

\begin{ex} \rm
The event poset $\cE$ in Figure~\ref{fig:ex1} has five linear extensions $\pi$:
\begin{eqnarray*}
\pi \quad \,\,\,& {\bf f}(\pi) \\
(1, 2, 3, 4) &  (1+f_{1000})(1+f_{1100})(1+f_{1110})  \\
(1, 2, 4, 3) & (1 + f_{1000}) (1+f_{1100}) f_{1101}         \\
(2, 1, 3, 4) & f_{0100}(1+f_{1100})(1+f_{1110})          \\
(2, 1, 4, 3) & f_{0100}(1+f_{1100}) f_{1101}                 \\
(2, 4, 1, 3) & (1+f_{0100}) f_{0101} (1+f_{1101})
\end{eqnarray*}
The sum of these five products equals the risk polynomial  $\RP(\cG;{\bf f}) $.
\end{ex}

\subsection*{Implementation}

Pruesse and Ruskey \cite{pruesse1994} showed that
the linear extensions of a poset $\cE$ can be computed in time linear in 
the number of linear extensions.
Thus, their algorithm computes $\RP(\cG;{\bf f}) $ in
time linear in the size of the output of
Theorem~\ref{thm:linearExtensions}.  That output is in
factored form (\ref{FPi}) and is always more compact than the
expanded risk polynomial.  In this manner, we compute the risk
polynomial in time sublinear in the size of the expanded risk
polynomial.  

To obtain the univariate risk polynomial, we take the sum of the terms
$\,(1+a)^{n-1-\delta} a^\delta$, where $\delta = \delta(\pi)$
is the number of descents of the linear extension $\pi$.
Similarly, the graded risk polynomial $\RP(\cG; a_1,\ldots,a_{n-1})$ is found by
keeping track of the descent set of each linear extension $\pi$.
We believe that this method is best possible for general posets
$\cE$. Notice that the leading term of
the univariate risk polynomial is the number of linear extensions of
$\cE$, and it is \#P-complete to count linear extensions \cite{brightwell}.  

When $\cE$ is a directed forest, the
recursive structure can be used to help compute the risk polynomial.
In this case, $\cE$ is built up by the operations of disjoint union
and ordinal sum from the one element poset.  For example, in the univariate case,
the zeta polynomial \cite[Sec.~3.11]{Stanley1999} of $\cG$ behaves nicely under these operations and
can be used to write down the risk polynomial. Based on these
considerations, we can design an efficient algorithm for
computing the univariate risk polynomial of a directed forest.

Using the method of Theorem~\ref{thm:linearExtensions}, we have developed software
for computing risk polynomials.
The input to our program is an arbitrary event poset $\cE$,
and the output is 
the risk polynomial, the graded risk polynomial
or the univariate risk polynomial.  Optionally, the user can also
input either exact fitness values or upper and lower bounds for each fitness
value.  The output in this case is either the exact risk of escape 
or upper and lower bounds for the risk.
It is designed to integrate with the package 
\texttt{Mtreemix} \cite{Beerenwinkel2005b},
allowing the user to start with data, infer a mutagenetic 
tree, and then easily compute the risk 
polynomial.
Our software is available at
\[ 
   \url{http://bio.math.berkeley.edu/riskpoly/}
\] 
We use the algorithm of \cite{Varol1981} for computing linear
extensions.  Although this algorithm isn't asymptotically optimal, as
shown in \cite{pruesse1994}, it
is simple to implement and efficient in practice.


\end{document}